\documentclass[12pt]{iopart}
\pdfoutput=1

\expandafter\let\csname equation*\endcsname\relax
\expandafter\let\csname endequation*\endcsname\relax 

\usepackage[english]{babel}
\usepackage{amsmath}
\usepackage{amssymb}
\usepackage{graphicx}
\usepackage{color}
\usepackage{graphicx}
\usepackage{caption}
\usepackage{subcaption}
\usepackage[utf8]{inputenc}
\usepackage{lmodern}
\usepackage[T1]{fontenc}
\usepackage{amsmath}
\usepackage{amsfonts}
\usepackage{graphicx} 
\usepackage{amsthm}
\usepackage{verbatim}
\usepackage{booktabs}
\usepackage{cite}
\usepackage{color}
\usepackage{bm}
\usepackage[colorlinks=true,linkcolor=red,citecolor=blue]{hyperref}

\usepackage[all]{xy}
\usepackage[width=17.5cm, vscale=0.8]{geometry}
\usepackage{booktabs}
\usepackage{libertine}

\def\Ord{\mathcal{O}}
\newcommand{\barr}{\begin{eqnarray}}
\newcommand{\earr}{\end{eqnarray}}
\newcommand{\beq}{\begin{equation}}
\newcommand{\eeq}{\end{equation}}
\newcommand{\be}{\begin{equation}}
\newcommand{\ee}{\end{equation}}

\newcommand{\de}{\mathrm{d}}

\newcommand{\avg}[1]{\left< #1 \right>} % for average
\let\baraccent=\= % rename builtin command \= to \baraccent
\renewcommand{\=}[1]{\stackrel{#1}{=}} % for putting numbers above =

{\left\lbrace\begin{array}{@{}l@{}}}%
{\end{array}\right.}

\newtheoremstyle{nopar}%
{}{}%
{\itshape}{}%
{\bfseries}{.\,\,}%
{ }%
{\thmname{#1}\thmnumber{ #2}: \thmnote{ #3}}
\theoremstyle{nopar}
\newtheorem{thm}{}
\newcounter{tmp}
\theoremstyle{plain}

\theoremstyle{remark}
\newtheorem{rmk}{Remark}

\newcommand{\kk}{\kappa}
\newcommand{\Ss}{\mathcal{S}}
\newcommand{\PP}{\mathcal{P}}

\newcommand{\V}{\mathrm{var}}
\newcommand{\Cov}{\mathrm{cov}}

 \newcommand{\EE}{\mathcal{E}}
\newcommand{\supp}{\operatorname{supp}}

\begin{document}

\title{Large deviations of spread measures for Gaussian matrices}

\author{Fabio Deelan Cunden$^{1}$ and Pierpaolo Vivo$^{2}$}
\address{$1.$ School of Mathematics, University of Bristol, University Walk, Bristol BS8 1TW, England\\
$2$. King's College London, Department of Mathematics, Strand, London WC2R 2LS, United Kingdom}

\date{\today}

\begin{abstract} 
For a  large $n\times m$ Gaussian matrix, we compute the joint statistics, including large deviation tails, of generalized and total variance - the scaled log-determinant $H$ and trace $T$ of the corresponding $n\times n$ covariance matrix. Using a Coulomb gas technique, we find that the Laplace transform of their joint distribution $\mathcal{P}_n(h,t)$ decays for large $n,m$ (with $c=m/n\geq 1$ fixed) as $\hat{\mathcal{P}}_n(s,w)\approx \exp\left(-\beta n^2 J(s,w)\right)$, where $\beta$ is the Dyson index of the ensemble and $J(s,w)$ is a $\beta$-independent large deviation function, which we compute exactly for any $c$. The corresponding large deviation functions in real space are worked out and checked with extensive numerical simulations. The results are complemented with a finite $n,m$ treatment based on the Laguerre-Selberg integral. The statistics of atypically small log-determinants is shown to be driven by the split-off of the \emph{smallest} eigenvalue, leading to an abrupt change in the large deviation speed.

\end{abstract}

%\pacs{}

\maketitle
\section{Introduction}\label{sec:intro}
The standard deviation $\sigma$ of an array of $m$ data $X_i$ is the simplest measure of how \emph{spread} these numbers are around their average value $\bar{X}=(1/m)\sum_{i=1}^m X_i$. Suppose that the $X_i$'s represent the final `Physics' marks of $m$ students of a high-school. Most worrisome scenarios for the headmaster would be a low $\bar{X}$ and/or a high $\sigma$, signaling an overall poor and/or highly non-uniform performance. 

What if `Physics' and `Arts' marks are collected together? Detecting performance issues now immediately becomes a much harder task, as data may fluctuate together and in different directions. A two-dimensional scatter plot may help, though. The ``centre" of the cloud gives a rough indication of how well the students perform on average in both subjects. But how to tell in which subject the gap between excellent and mediocre students is more pronounced, or whether outstanding students in one subject also excel in the other? 

\begin{figure}[t]
\centering
\includegraphics[width=.65\columnwidth]{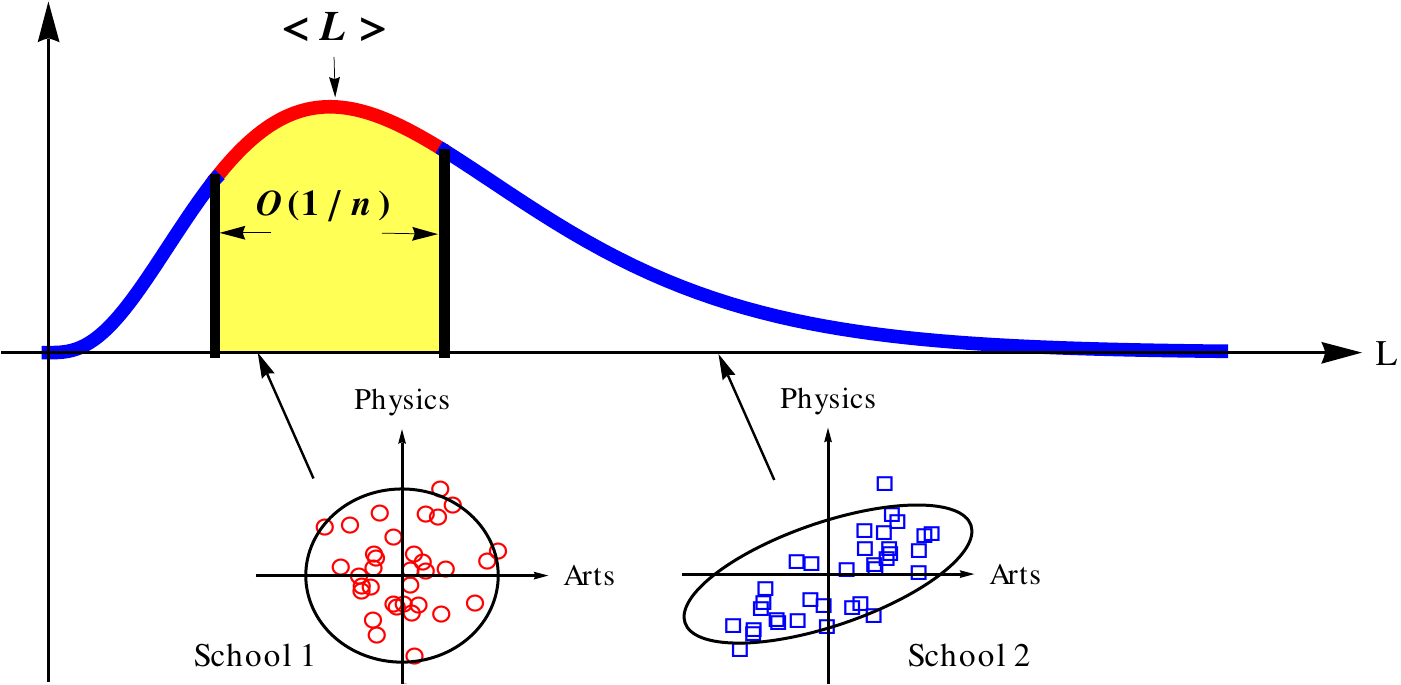}
\caption{(color online) Top: Sketch of the probability density of the likelihood ratio $L$ of a Gaussian iid data set. In yellow, the typical region around the mean of order $\mathcal{O}(1/n)$. Larger fluctuations are referred to as atypical large deviations.  Bottom: Sketch of two multivariate data sets with $n=2$ and $m=35$. Each point represents a student, for two different schools, and his/her marks in Arts and Physics. 
The two datasets have same generalized variance $H$, but different total variance $T$. The likelihood ratio $L$ of School 1 is compatible with the iid hypothesis, while the value of $L$ for School $2$ is atypically far from the average $\avg{L}$.}
\label{fig:scatter}
\end{figure}

In Fig. \ref{fig:scatter} (Bottom) we sketch two scatter plots of marks adjusted to have zero mean. A meaningful spread indicator seems to be the shape of the ellipse enclosing each cloud. For example, an almost circular cloud - like School $1$ - represents a rather uninformative situation, where your `Arts' marks tell nothing about your `Physics' skills, and vice versa. Conversely, a rather elongated shape - like School $2$ - highlights correlations between each student's marks in different subjects.

For a bunch of many scattered points it would be desirable to summarize the overall spread around the mean just by a single scalar quantity, like the \emph{perimeter} or \emph{area} of the enclosing ellipse. Not surprisingly, however, these indicators (taken individually) have evident shortcomings \cite{Wichern}. Surely a wiser choice is to combine more than a single spread measure (like perimeter \emph{or} area \emph{alone}), to obtain a more revealing indicator.
These issues arise naturally in multivariate statistics, and more mathematical tools and techniques are required compared to the univariate setting. 

In this work, we compute the \emph{joint} statistics of ``perimeter" and ``area" enclosing clouds of random high-dimensional data. Why this is a crucial (and so far unavailable) ingredient for an accurate data analysis will become clearer very shortly. 

In the more general setting of $n$ subjects and $m$ students, their marks can be arranged in a $n\times m$ matrix $\mathcal{X}$, adjusted to have zero-mean rows. We then construct the normalized $n\times n$ covariance data matrix  $\Ss=(1/n)\mathcal{X} \mathcal{X}^{\dagger}$, with non-negative eigenvalues $(\lambda_1,\dots,\lambda_n)$, which is precisely the multi-dimensional analogue of the variance $\sigma^2$ for a single array. The surface and volume (``perimeter'' and ``area'' in the two-dimensional example) of the enclosing ellipsoid are related to the scaled trace and determinant of $\mathcal{S}$:
\be
T=\frac{1}{n}\Tr\Ss\quad\mathrm{and}\quad
G=\det\Ss^{1/n} .
\ee
 In statistics, these objects are called \emph{total} and \emph{generalized} variance respectively \cite{anderson}. 
As discussed before, blending both estimators together would be preferable, like in the widely used positive scalar combination 
\be
L = T-H-1\ ,
\ee 
called \emph{likelihood ratio} \cite{anderson}, where $H$ is the log-determinant of $\Ss$
\be
H=\ln G=\frac{1}{n}\Tr\ln\Ss\ .
\ee
Values of $L$ for different shapes of the data cloud are sketched in Fig.~\ref{fig:scatter} (Bottom).

Now, suppose that we wish to test the hypothesis that the data $X_{ij}$ (yielding a certain \emph{empirical} $L$) are independent and identically distributed. 
What if an \emph{atypically} high or low $L$ (with respect to a null i.i.d. model) comes out from the data? We would be tempted to reject the test hypothesis outright. However, this might lead to a misjudgment, as atypical values of $L$ for the null model \emph{can} (and \emph{do}) occur (just very rarely). 
What is the probability of this rare event? Here we provide a solution to this problem, computing the joint statistics of total and generalized variance for a large Gaussian dataset. 

The derivation of these results relies on techniques borrowed from statistical mechanics and random matrix theory (RMT). We express the large deviation functions of spread indicators as excess free energies of an associated 2D Coulomb gas, whose thermodynamic limit is analyzed in the mean-field approximation valid for $n,m\to\infty$ with $m/n>1$ fixed. This approach is complemented with a finite $n,m$ analysis based on the ``Laguerre" version of the celebrated Selberg integral. The marriage between these two techniques provides an elegant solution to a challenging problem. In addition, our unifying framework recovers and extends some partial results earned by statisticians via other techniques.

The article is organized as follows. In Section~\ref{sec:results} we introduce the notation and we summarize our main results. We then elaborate at length on its consequences. Finally, we briefly discuss the relation of our findings with earlier works. Section~\ref{sec:derivation} contains the derivations. First we summarize the ``Coulomb gas method'' and we present a quite general algorithm to find the large deviation functions of linear statistics on random matrices (Subsection~\ref{sec:coulomb}). Then, in Subsection~\ref{sec:LDP} we turn to the actual proof. In Section~\ref{sec:finite} we discuss two issues that are not captured by the Coulomb gas method. Finally we conclude with a summary and some open questions in Section~\ref{sec:conclusions}.

\section{Setting and formulation of the results} 
\label{sec:results}
We consider an ensemble of $n\times m$ matrices $\mathcal{X}$ whose entries are real, complex or quaternion independent standard Gaussian variables \footnote{The assumption of independence is not restrictive. If the entries of $\mathcal{X}$ are centered correlated Gaussian variables with positive definite covariance matrix $\Sigma$ our methods can be applied to the matrix $\Sigma^{-1/2}\mathcal{S}\Sigma^{-1/2}$.}, labeled by Dyson's index $\beta=1,2$ and $4$ respectively, and we form the $n\times n$ (real, complex or quaternion) sample covariance matrix
\be
\mathcal{S}=\frac{1}{n}\,\mathcal{X} \mathcal{X}^\dagger\ .\label{eq:covariance_matrix}
\ee
This ensemble of random covariance matrices (positive semi-definite by construction) is known as the Wishart ensemble~\cite{Wishart} with rectangularity parameter $c=m/n\geq 1$. Remarkably, in the Gaussian case, the joint probability density $\PP(\lambda_1,\dots,\lambda_n)$ of the positive $\mathcal{O}(1)$ eigenvalues of $\mathcal{S}$ is known explicitly~\cite{fisher,anderson}
\begin{align}
\PP(\lambda_1,\dots,\lambda_n)&=\frac{1}{\mathcal{Z}_n} \e^{-\beta E[\bm\lambda]},\quad E[\bm\lambda]=-\frac{1}{2}\sum_{i\neq j}\ln{|\lambda_i-\lambda_j|}+n\sum_{k}V(\lambda_k)\ ,\label{eq:jpd_spread}
\end{align}
where the \emph{energy} function $E[\bm\lambda]$ contains the external potential 
\be
V(\lambda)=
\begin{cases}
 \displaystyle \frac{\lambda}{2}-\alpha\ln \lambda  &\text{for $\lambda>0$ if $\alpha>0$ \,(or $\lambda\geq0$ if $\alpha=0$)}\\
+\infty\ &\text{otherwise}
\end{cases}\label{eq:potential_param}\quad\text{with}\,\,\,\alpha=\frac{c-1}{2}+\frac{1}{2n}-\frac{1}{\beta n}\ .
\ee
The normalization constant $\mathcal{Z}_n=\int  \e^{-\beta E[\bm\lambda]}\de\bm\lambda$ is also known for any finite $n$ from the celebrated Selberg integral~\cite{Selberg42,forrester,Guionnet}.
 The joint law of the eigenvalues \eqref{eq:jpd_spread} is the Gibbs-Boltzmann canonical distribution of a 2D Coulomb gas (logarithmic repulsion) constrained to stay on the positive half-line and subject to the external potential $V$ at inverse temperature $\beta$ (we adopt the usual physical convention that probabilities are zero in regions of infinite energy). As we shall see, the derivation of our result is independent of the restriction $\beta=1,2$ or $4$. Therefore, from now on we shall consider non-quantized values\footnote{Eigenvalues obeying the Wishart statistics with general $\beta>0$ can be generated efficiently using Dumitru-Edelman tridiagonal construction~\cite{Dumitriu02}.} $\beta>0$.

We consider the scaled log-determinant $H$ and trace $T$ of the covariance matrix $\Ss$. In terms of the eigenvalues $\lambda_1,\dots,\lambda_n$, they read
\be
H=\frac{1}{n}\sum_{i=1}^n \ln\lambda_i \quad\mathrm{and}\quad
T=\frac{1}{n}\sum_{i=1}^n \lambda_i.\label{eq:HT}
\ee
Their joint probability law and Laplace transform are denoted respectively by
\be
\mathcal{P}_n(h,t)=\avg{\delta\left(h-H\right)\delta\left(t-T\right)}\ ,
\quad\widehat{\mathcal{P}}_n(s,w)=\avg{\e^{-\beta n^2(sH+wT)}},
\label{eq:def}
\ee
where the average is taken with respect to the canonical distribution of the eigenvalues~\eqref{eq:jpd_spread}. Here we are interested in the large $n$ behavior of $\mathcal{P}_n(h,t)$ and $\widehat{\mathcal{P}}_n(s,w)$ at logarithmic scales. More precisely, we show that for large $n$
\be
\mathcal{P}_n(h,t)\approx  \e^{-\beta n^2 \Psi(h,t)}\qquad \text{and}\qquad\widehat{\mathcal{P}}_n(s,w)\approx  \e^{-\beta n^2 J(s,w)}\ ,
\label{eq:LDP}
\ee
where $a_n\approx b_n$ stands for $\ln a_n/\ln b_n\to1$ as $n\to\infty$.

The functions $\Psi(h,t)$ and $J(s,w)$ are called \emph{rate function} and \emph{cumulant generating function} (GF) respectively~\cite{Ellis84,Touchette09}. It is a standard result in large deviation theory that the functions $\Psi(h,t)$ and $J(s,w)$ in~\eqref{eq:LDP} are related via a Legendre-Fenchel transformation.

Here we compute explicitly, for all $\beta>0$ and $c=m/n\geq 1$, the cumulant GF
\be
J(s,w)=-\lim_{n\to\infty}\frac{1}{\beta n^2}\ln \widehat{\mathcal{P}}_n(s,w).  \label{eq:GF_spread2}
\ee 
From now on we shall denote $s_c=(c-1)/2$. The main results of the paper are as follows.
\smallskip

\begin{thm}[{Joint large deviation function of Generalized and Total Variances}]\label{thm:A}

Let $(\lambda_1,\dots,\lambda_n)$ be distributed according to \eqref{eq:jpd_spread}-\eqref{eq:potential_param} and let $H$ and  $T$ as in~\eqref{eq:HT}. Their joint cumulant generating function $J(s,w)$ defined by \eqref{eq:GF_spread2} exists for $s\leq s_c$ and $w>-1/2$ 
and is given by
\be
J(s,w)=J_H(s)+J_T(w)-s\ln{\left(1+2w\right)}\ ,
\label{eq:jointJxy}
\ee
where $J_T(w)$ and $J_H(s)$ are the individual GF of cumulants of  $T$ and $H$. They are given explicitly by
\begin{align}
J_H(s)&=-\lim_{n\to\infty}\frac{1}{\beta n^2}\ln\widehat{\mathcal{P}}_n(s,0)=\phi\left(s-s_c\right)-\phi\left(-s_c\right)\ ,\label{eq:detSCGF_thm}\\
J_T(w)&=-\lim_{n\to\infty}\frac{1}{\beta n^2}\ln\widehat{\mathcal{P}}_n(0,w)=\frac{c}{2}\ln{\left(1+2w\right)}\ , \label{eq:traceSCGF_thm}
\end{align}
with $\phi(x) =-\frac{3}{2}x+x^2\ln (-2x)-\frac{(1-2x)^2}{4}\ln{\left(1-2x\right)}$ for $x\leq0$.
\end{thm}
We first discuss some consequences this result. The derivation is postponed to Section~\ref{sec:derivation}.

\begin{rmk} The large deviation functions $J(s,w)$, $J_H(s)$ and $J_T(w)$ are independent of $\beta$. This property is standard for 2D Coulomb gas systems. 
\end{rmk}
\begin{rmk}
The joint cumulant GF is not the sum of the single generating functions: $J(s,w)\neq J_H(s)+J_T(w)$ ($T$ and $H$ are not independent for large $n$). 
\end{rmk}
\begin{rmk} For $c>1$ the GF is analytic at $s=w=0$ and the joint cumulants of $T$ and $H$ are obtained by evaluating the derivatives of $J(s,w)$ at $(s,w)=(0,0)$. More precisely, to leading order in $n$ for $\kk,\ell\geq0$  
\be
C_{\kk,\ell}(H,T)=\left(-\beta n^2\right)^{1-(\kk+\ell)}\frac{\partial^{\,\kk+\ell} }{\partial^{\kk}s\,\partial^{\ell}w}J(s,w)|_{s=w=0}\ .
\label{eq:cumul}
\ee
Note that $J(0,0)=0$. Extracting the first cumulants, we obtain 
to leading order in $n$
\begin{align}
&\avg{T}=c , \qquad \avg{H}=-1-(c-1) \ln(c-1)+c \ln c\ , \label{averageTH}\\
&\frac{\V(T)}{\omega_{\beta}(n,2)}=c,\quad \frac{\V(H)}{\omega_{\beta}(n,2)}=\ln{\frac{c}{c - 1}} ,\quad\frac{\Cov(T,H)}{\omega_{\beta}(n,2)}=1\ ,\label{varTH}
\end{align}
where we set $\omega_{\beta}(n,\ell)=\left(2/\beta n^2\right)^{\ell-1}$. The correlation coefficient
$(\Cov(H,T)/\sqrt{\V(T)\V(H)})=1/\sqrt{c\ln(c/(c-1))}$,
independent of $\beta$, is positive for all values of $c$ (if the ``area'' increases, typically so does the ``perimeter''). Notice that the expression of $\V(H)$ does not cover the case $c=1$ (square data matrices). This case will be treated separately in Section~\ref{sec:finite}.

The decay of the higher order mixed cumulants $C_{\kappa,\ell}(H,T)$ for $\kappa+\ell>2$ is given to leading order in $n$ by
\be
C_{\kappa,\ell}(H,T)=\omega_{\beta}(n,\kappa+\ell)\left\{(\kappa-3)!\left[(1-c)^{2-\kappa}-(-c)^{2-\kappa}\right]\delta_{\ell,0}+c\ (\ell-1)!\delta_{\kappa,0}+(\ell-1)!\delta_{\kappa,1}\right\}\ .\label{eq:highc}
\ee

\end{rmk}
\begin{rmk}\label{rmk:4} The marginal  probability densities  $\PP_H(h)=\avg{\delta(h-H)}$ and $\PP_T(t)=\avg{\delta(t-T)}$  behave as
\be
\PP_H(h)\approx \e^{-\beta n^2\Psi_H(h)}\,\quad \text{and}\quad \PP_T(t)\approx \e^{-\beta n^2\Psi_T(t)},\label{beh}
\ee
where $\Psi_H(h)$ and $\Psi_T(t)$ are the individual rate functions of $T$ and $H$. These individual rate functions should also be in principle computable as inverse Legendre-Fenchel transform of~\eqref{eq:detSCGF_thm}-\eqref{eq:traceSCGF_thm}. However, for the scaled log-determinant $H$ this is only possible for ``not too small" values ($h>-1$); this point will be discussed in more details in Section~\ref{sec:finite}. 
 The expression of $\Psi_T(t)$ in the full range is instead remarkably simple
\be
\Psi_T(t)=\frac{t-c}{2}+\frac{c}{2}\ln{\left(\frac{c}{t}\right)}\ ,\qquad\quad  (t>0)\ .
\label{eq:traceratefunct}
\ee
This analytic function is strictly convex and positive and it attains its unique minimum (zero) at $t= c$ (the asymptotic mean value of $T$, see~\eqref{averageTH}). This rate function provides information on the large $n$ full probability density of $T$. We can identify three regimes: 
\begin{itemize}
\item[i)] typical fluctuations of order $\mathcal{O}(1/n)$ about the average are described by the quadratic behavior of $\Psi_T(t)$ around its minimum at $t=c$, corresponding to  asymptotically Gaussian fluctuations with mean and variance as in~\eqref{averageTH}-\eqref{varTH};
\item[ii)] large deviations for $t\gg c$ (atypically \emph{large} ``perimeters'') exhibit an exponential decay (independent of the rectangularity parameter $c$);
\item[iii)] for $t\ll c$ (atypically \emph{small} ``perimeters'') we find a $c$-dependent power law. 
\end{itemize} 
Summarizing:
\begin{equation}
\mathcal{P}_T(t)\approx  \e^{-\beta n^2\Psi_T(t)}
\sim
\begin{cases}
 t^{\beta n^2 c/2}\ ,&\quad (t\to 0)\ \\
  \e^{-\beta n^2 \frac{(t-c)^2}{4c}}\ ,&\quad (t\sim c)\ \\
  \e^{-\beta n^2 t/2}\ ,&\quad (t\to +\infty)\ .
\end{cases}
\end{equation}
These predictions have been confirmed by extensive numerical simulations. A sample size of about $N=10^8$ spectra of complex ($\beta=2$) Wishart matrices has been efficiently generated using a tridiagonal construction~\cite{Silverstein85}. The data are plotted in Fig.~\ref{fig:ratespread} and show a very good agreement with the behavior in \eqref{beh}.
\begin{figure}[t]
\centering
\includegraphics[width=.99\textwidth]{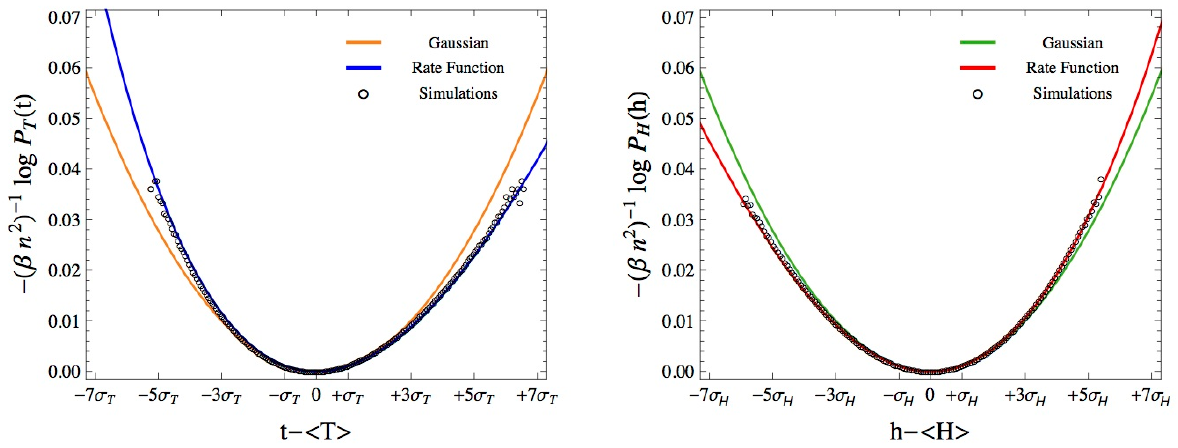}
\caption{
Numerical simulations (black circles)   of complex ($\beta=2$) Wishart matrices $\Ss$ of size $n=15$ with $c=2$. Here the sample size is $N=2.5\cdot10^8$.
Left: The numerical values (black circles) for the total variance $T=n^{-1}\sum_{i}\lambda_i$. The Gaussian approximation (orange line) with average $\avg{T}$~\eqref{averageTH} and standard deviation $\sigma_T=\sqrt{\V(T)}$ \eqref{varTH} fits well the data for fluctuations of order $\sim 3\sigma_T$ but deviates strongly for atypical fluctuations. The global behavior is captured instead by the large deviation function (blue line) $\Psi_T(t)$ of~\eqref{eq:traceratefunct}. Right: Numerical values (black circles) for the log-determinant $H=n^{-1}\sum_{i}\ln\lambda_i$. Again, the Gaussian approximation (green line) with average $\avg{H}$~\eqref{averageTH} and standard deviation $\sigma_H=\sqrt{\V(H)}$ \eqref{varTH} describes well the data for fluctuations of order $\sim 3\sigma_H$ but deviates for larger fluctuations. The large deviation function (red line) $\Psi_H(h)$ of~\eqref{eq:rateH} provides a global description of the data. The critical point $h=-1$ (below which the large deviations change speed from $n^2$ to $n$) is not visible in the picture (for $n=15$  and $c=2$ the critical point is at $\sim 25\sigma_H$ to the left of $\avg{H}$). 
}
\label{fig:ratespread}
\end{figure}

\begin{comment}
\begin{figure}[t]
\centering
\includegraphics[width=.6\textwidth]{spread.pdf}
\caption{Numerical simulations (black circles) for the total variance $T=n^{-1}\sum_{i}\lambda_i$  of complex ($\beta=2$) Wishart matrices $\Ss$ of size $n=15$. Here the sample size is $N=2.5\cdot10^8$. The Gaussian approximation (orange line) with average $\avg{T}$~\eqref{averageTH} and standard deviation $\sigma_T=\sqrt{\V(T)}$ \eqref{varTH} fits well the data for fluctuations of order $\sim 3\sigma_T$ but deviates strongly for atypical fluctuations. The global behavior is captured instead by the large deviation function (blue line) $\Psi_T(t)$ of~\eqref{eq:traceratefunct}.}
\label{fig:ratespread}
\end{figure}
\end{comment}
\end{rmk}

Once the joint large-$n$ behavior of generalized and total variances is known, one may easily derive a large deviation principle for any continuous function of them. For instance, from $\widehat{\mathcal{P}}(s,w)$, it is easy to compute the Laplace transform of the likelihood ratio $L=T-H-1$ as $\widehat{\mathcal{P}}_L(s)=\langle \e^{-\beta n^2 sL}\rangle
= \e^{\beta n^2 s}\widehat{\mathcal{P}}(-s,s)$. Hence we have the following result.  
\smallskip

\begin{thm}[Large deviations of the likelihood ratio]\label{cor:1} The likelihood ratio cumulant GF is given by
\be
J_L(s)=-\lim_{n\to\infty}\frac{1}{\beta n^2}\ln\widehat{\mathcal{P}}_L(s)=J(-s,s)-s\qquad \text{for}\,\,\, -1/2<s\leq s_c\ ,\label{eq:lapl_L}
\ee
with $J$ as in~\eqref{eq:jointJxy}.
With the same notation as above, the cumulants of $L$ at leading order in $n$ follow by differentiations 
\be
C_\ell(L)=C_\ell(T)+(-1)^{\ell}C_\ell(H)+\delta C_{\ell}\quad \mathrm{with}\quad \delta C_\ell=\omega_{\beta}(n,\ell)\frac{\ell\,!}{(1-\ell)}\theta\left(\ell-1\right)-\delta_{\ell,1}\ ,\label{eq:cumulL}
\ee
($\theta$ is the Heaviside step function) for $\ell\geq1$. This corresponds to typical fluctuations on a region $\Ord(1/n)$ around the mean
\be
\avg{L}=c+(c-1)\ln{(c-1)}-c\ln{c} ,
\ee 
with variance\footnote{Again, these results are not valid for $c=1$.}  
\be
\V(L)=\omega_{\beta}(n,2)\left[c+\ln{(c/(c-1))}-2\right]. \label{varL}
\ee 
\end{thm}
Note that, since $T$ and $H$ are not independent, the cumulants~\eqref{eq:cumulL} of $L$ involve the extra term $\delta C_{\ell}$.

From Result~\ref{thm:A}, extracting the asymptotics of the first moments of $T$ and $H$ for $c\gg1$ we recover classical results in multivariate analysis, valid when the sample size $m$ is much larger than the number of variates $n$. 
\smallskip

\begin{thm}[Classical statistics]\label{cor:2} In the regime $m\gg n\gg 1$, $T$ and $H$ become asymptotically Gaussian. More precisely, as $c\to\infty$ 
\begin{align}
\sqrt{\frac{\beta n^2}{2c}}\left(T-c\right)\to\mathcal{N}(0,1), \qquad\text{and}\qquad
\sqrt{\frac{\beta c n^2}{2}}\left(H-\ln{c}\right)\to\mathcal{N}(0,1) \label{asymTH2},
\end{align}
in distribution, where $\mathcal{N}(0,1)$ denotes a standard Gaussian variable. 
\end{thm}

To conclude this introductory section, we remark that our findings reproduce some known results for the typical fluctuations (mean and variance) of $T, H$ and $L$ separately, in the real case ($\beta=1$) \cite{anderson,Johnstone,Bai,jonsson,jiang}. Moreover, the variances and covariances~\eqref{varTH} and~\eqref{varL} can be computed for generic $\beta$  using covariance formulae valid for one-cut $\beta$-ensembles of random matrices~\cite{Beenakker94,Cunden14}.

A precious tool in classical statistics is the \emph{Barlett decomposition}~\cite{Barlett33}, which is useful to transform functions of strongly correlated eigenvalues of Wishart matrices (see~\eqref{eq:jpd_spread}) into functions of independent (but not identical) chi-squared random variables. In the asymptotic regime $m\gg n$ this decomposition becomes sufficiently manageable to derive some interesting results. For real matrices, the limits~\eqref{asymTH2} agree with classical theorems based on the Barlett decomposition (see e.g.~\cite{Muirhead}). From the results on $H$, the statistical behavior of the scaled determinant $G=\e^H$ can be easily derived. For statistics of determinants of random matrices (more general than the sample covariance matrices considered here), see~\cite{Cicuta,Rouault05,caer1,Tao12}. For more details on the classical methods in multivariate analysis we refer to the classical books~\cite{Muirhead,Wichern} and the excellent review~\cite{Johnstone} on the applications of RMT in multivariate statistics.

\section{Derivation}\label{sec:derivation}
We now turn to the derivation of Result~\ref{thm:A}. Results~\ref{cor:1} and~\ref{cor:2} follow as corollaries and hence their proof will be omitted. In Subsection~\ref{sec:coulomb} we set up the variational problem in the framework of the 2D Coulomb gas thermodynamics. The Coulomb gas analogy for spectra of random matrix ensembles goes back to the seminal works by Wigner~\cite{Wigner57} and Dyson~\cite{Dyson62}. In particular, it was Dyson who first used this analogy to compute large random matrix statistics. 
This idea has been developed later and used in several areas of physics~\cite{Brezin78,Chen94,Dean06,Vivo07,Facchi08,DePasquale10,Majumdar11,Vivo12,Facchi13,Texier13,Majumdar13,
Majumdar14,Grabsch15,Cunden15,Cunden15b,Colomo13,Cunden_unp,CundenMV15}.
In Subsection~\ref{sec:LDP} we solve the saddle-point equations and we compute explicitly $J(s,w)$, thus proving Result~\ref{thm:A}.

\subsection{2D Coulomb gas problem}\label{sec:coulomb}
The Coulomb gas calculation goes as follows. First we observe from~\eqref{eq:jpd_spread}-\eqref{eq:potential_param} that 
the joint Laplace transform~\eqref{eq:def} is finite for $s\leq \alpha=s_c+\Ord\left(n^{-1}\right)$ and $w>-1/2$.
From~\eqref{eq:jpd_spread} and~\eqref{eq:HT}-\eqref{eq:def}, this Laplace transform 
can be written as the ratio of two partition functions
\begin{align}
\widehat{\mathcal{P}}_n(s,w)&= \left[\mathcal{Z}_n(s,w)/\mathcal{Z}_n(0,0)\right] \label{laplsw_ratio},\\
\mathcal{Z}_n(s,w)&=\int\de\lambda_1\cdots\de\lambda_n\,  \e^{-\beta E[\bm\lambda;s,w]}\ .\label{laplsw}
\end{align}
$\mathcal{Z}_n(s,w)$ is the partition function of a constrained Coulomb gas, where the energy function $E[\bm\lambda;s,w]=E[\bm\lambda]+\sum_k U_{s,w}(\lambda_k)$ contains now the additional single-particle potential
\be
U_{s,w}(\lambda)=s\ln\lambda+w\lambda\ .
\ee
Note that $\mathcal{Z}_n(0,0)$ is the partition function of the unconstrained gas and therefore it coincides with the normalization constant in~\eqref{eq:jpd_spread}.

Hence, the computation of the joint cumulant GF $J(s,w)$ amounts to evaluating the leading order in $n$ of the partition function $\mathcal{Z}_n(s,w)$. More precisely, from~\eqref{laplsw_ratio}-\eqref{laplsw}, $J(s,w)$ may be expressed as the \emph{excess free energy}
\be
J(s,w)=-\lim_{n\to\infty}\frac{1}{\beta n^2}\left[\ln\mathcal{Z}_n(s,w)-\ln\mathcal{Z}_n(0,0)\right]
\label{eq:excess}
\ee
of the Coulomb gas in the \emph{effective potential} $V(\lambda)+U_{s,w}(\lambda)$ with respect to the unperturbed Coulomb gas ($s=w=0$). This effective potential is bounded from below for $s\leq \alpha=s_c+\Ord\left(n^{-1}\right)$ and $w>-1/2$ (the domain of existence of the Laplace transform $\widehat{\mathcal{P}}_n(s,w)$, of course). For any finite $n,m$, the excess free energy~\eqref{eq:excess} can be computed exactly in terms of a Laguerre-Selberg integral~\cite{Selberg42} (see Section~\ref{sec:finite}). How to deal with the limit of large $n$? We show now how the task of computing $J(s,w)$ can be reduced to a variational problem.

First, we introduce the normalized density of the gas particles $\rho_n(\lambda)=n^{-1}\sum_{i=1}^n\delta(\lambda-\lambda_i)$, in terms of which any sum function on the eigenvalues $\lambda_1,\dots,\lambda_n$ can be easily expressed. For instance, the log-determinant and trace~\eqref{eq:HT}, both linear statistics on $\Ss$, are conveniently expressed as linear functionals on $\rho_n(\lambda)$ as
\be
H=\int\de\rho_n(\lambda)\,\ln\lambda\qquad \text{and}\qquad T=\int\de\rho_n(\lambda)\,\lambda\ .\label{HTint}
\ee 

Second, for large $n$, the energy function $E[\bm\lambda;s,w]$ of the 2D Coulomb gas can be converted into a \emph{mean-field energy functional} $E[\bm\lambda;s,w]\sim n^2\mathcal{E}[\rho_n;s,w]$,
where
\be
\mathcal{E}[\rho;s,w]=-\frac{1}{2}\iint_{\lambda\neq\lambda'}\!\!\!\!\!\de\rho(\lambda)\,\de\rho(\lambda')\,\ln|\lambda-\lambda'|+\int\de\rho(\lambda)\,V(\lambda)+\int\de\rho(\lambda)\,U_{s,w}(\lambda)\ .
\label{eq:mf_func}
\ee
The mean-field functional~\eqref{eq:mf_func} has been intensely studied in several fields. 
We refer to~\cite{Saff,Serfaty14,Cunden_unp} for a detailed exposition and collection of known results.
In particular, it is known that for large $n$ the partition function $\mathcal{Z}_n(s,w)$ is  
 dominated  by $\rho_{s,w}(\lambda)$, the unique minimizer of the mean-field energy functional
$\mathcal{E}[\rho;s,w]$ in the space of normalized densities:
\be
\mathcal{Z}_n(s,w)\approx \exp\left({-\beta n^2 \mathcal{E}[\rho_{s,w};s,w]}\right)\;\;\;\text{with}\;\;\;\mathcal{E}[\rho_{s,w};s,w]=\min_{\substack{\rho\geq0\\\int\!\de\rho=1}}\mathcal{E}[\rho;s,w]\ .
\label{eq:saddle-point_prob}
\ee
The meaning of the saddle-point density is the following: $\rho_{s,w}(\lambda)$ is the typical configuration of the eigenvalues yielding 
a prescribed value of log-determinant and trace 
\be
h(s,w)=\int\de\rho_{s,w}(\lambda)\,\ln\lambda\qquad t(s,w)=\int\de\rho_{s,w}(\lambda)\,\lambda\ .\label{eq:ht}
\ee
Hence, a possible route to evaluate $J(s,w)$ consists of finding for all $s,w$ the saddle-point density $\rho_{s,w}(\lambda)$ and inserting it back into the energy functional~\eqref{eq:mf_func} to evaluate the leading order of $\mathcal{Z}_n(s,w)$ as in~\eqref{eq:saddle-point_prob}. This technique has been exploited in the last decade in many physical problems, mainly to compute the large deviations of \emph{single} observables. However, this route entails the explicit computation of the mean-field energy~\eqref{eq:mf_func} at the saddle-point density, which is not necessarily an easy task. The situation gets even worse in the case of joint statistics.

In certain situations one can use a shortcut (see~\cite{Cunden_unp}) based on a thermodynamic identity that has been stated rigorously in the language of large deviation theory~\cite{Touchette09} by G\"artner~\cite{Gartner77} and Ellis~\cite{Ellis84}. It is known that, if a cumulant GF  $J(\vec{s})$ is differentiable in the interior of its domain, then the rate function $\Psi(\vec{x})$ is the Legendre-Fenchel transform of the cumulant GF  (and hence, $J(\vec{s})$ is the inverse Legendre-Fenchel transform of $\Psi(\vec{x})$). This relation between rate function and cumulant GF can be exploited in our problem as follows (for a general mathematical discussion we refer to~\cite{Cunden_unp}). We assume first that $J(s,w)$ is differentiable. Therefore, the G\"artner-Ellis theorem ensures that $\Psi(h,t)$ is also smooth and given from $J(s,w)$ by the Legendre-Fenchel transformation
\be
\Psi(h,t)=\sup_{s,w}\left[J(s,w)-(sh+wt)\right]\ .\label{eq:Legendre}
\ee
The identity  \eqref{eq:Legendre} can be written in the (almost) symmetric form
\be
J(s,w)-\Psi(h,t)=sh+wt\ . \label{eq:sx}
\ee
This equation should be interpreted with care. Indeed, in \eqref{eq:sx}, there are only \emph{two} independent variables, for instance $s$ and $w$ \emph{or} $h$ and $t$. The relation between the conjugate variables $(h,t)$ and $(s,w)$ is provided by 
\be
\frac{\partial J(s,w)}{\partial s}=h(s,w)\,,\frac{\partial J(s,w)}{\partial w}=t(s,w)\quad\text{or equivalently}\quad
\frac{\partial \Psi(h,t)}{\partial h}=s(h,t)\,,\frac{\partial \Psi(h,t)}{\partial h}=w(h,t)\ ,
\ee
where $h(s,w)$ and $t(s,w)$ are given in~\eqref{eq:ht} and $s(h,t)$, $w(h,t)$ are the corresponding inverse maps. Hence, we can write the differential relations
\begin{align}
\de J(s,w)&=h(s,w)\de s+t(s,w)\de w\label{eq:diff1}\\
-\de \Psi(h,t)&=s(h,t)\de h+w(h,t)\de t\ ,  \label{eq:diff2}
\end{align}
supplemented with the normalization condition $J(0,0)=0$ (and hence $\Psi(h(0,0),t(0,0))=0$). The expressions~\eqref{eq:diff1} and~\eqref{eq:diff2} can be interpreted as 
Maxwell relations among thermodynamic potentials, in our case the Helmholtz free energy and the enthalpy. However it is somewhat astonishing that these relations have not been applied in the Coulomb gas computations until very recently (for applications of~\eqref{eq:diff1}-\eqref{eq:diff2} in physical models see~\cite{Cunden15,Cunden15b} and also~\cite{Grabsch15}). 

Using~\eqref{eq:diff1}-\eqref{eq:diff2} one can use the following shortcut to compute the large deviations functions (for a detailed exposition we refer to~\cite{Cunden_unp}). For instance, in order to compute $J(s,w)$ we only need to find the saddle-point density $\rho_{s,w}(\lambda)$ and compute $h(s,w)$ and $t(s,w)$ from~\eqref{eq:ht}. Then, $J(s,w)$ follows from integration of~\eqref{eq:diff1}
\be
J(s,w)=\int_{(0,0)}^{(s,w)}\!\de J(s,w)\ . \label{eq:int_diff}
\ee
This shortened route of the Coulomb gas method provides an effective tool to evaluate large deviations functions. A large amount of unnecessary computations can be avoided and the task of computing \emph{joint} large deviations becomes feasible. In the next subsection, we will use this strategy to derive our main result. 
\subsection{Saddle-point equation and large deviation functions}\label{sec:LDP}
The first problem to overcome is to find the saddle-point density $\rho_{s,w}$ of the mean-field functional $\EE[\rho;s,w]$. From~\eqref{eq:mf_func}, the stationarity condition of $\rho_{s,w}$ reads 
\be
\int\de\rho_{s,w}(\lambda')\,\ln|\lambda-\lambda'|-V(\lambda)-U_{s,w}(\lambda)=\mathrm{const},\, \qquad\text{for}\,\,\lambda\in\supp\rho_{s,w},
\label{eq:saddle}
\ee
where $\supp\rho_{s,w}$ denote the support of $\rho_{s,w}$ (for $\lambda\notin\supp\rho_{s,w}$ the left hand side is greater than or equal to the same constant). The physical meaning of~\eqref{eq:saddle} is clear: at equilibrium, the 2D Coulomb gas arranges itself in such a way that each particle  has equal electrostatic energy (the left hand side of~\eqref{eq:saddle}). 

Taking one further derivative with respect to $\lambda$, the resulting singular integral equation can be solved for $\rho_{s,w}(\lambda)$ using a theorem due to Tricomi (\cite[Sec.~4.3]{Tricomi57}), and the result reads
\be
\rho_{s,w}(\lambda)=\frac{1+2w}{2\pi\lambda}\sqrt{\left(\lambda-\lambda_{-}\right)\left(\lambda_{+}-\lambda\right)}\,{\bm 1}_{\lambda\in(\lambda_-,\lambda_+)} ,\label{rhostar}
\ee
where the edges $\lambda_{\pm}$ of the support depend on $s$ and $w$ as 
\be
\lambda_{\pm}(s,w)=\frac{1}{1+2w}\left(1\pm\sqrt{1-2\left(s-s_c\right)}\right)^2. \label{eq:support}
\ee
For $s=w=0$ the density of the unperturbed gas $\rho_{0,0}(\lambda)$ coincides  with the Mar\v{c}enko-Pastur distribution~\cite{MP} with edges $(1\pm\sqrt{c})^2$. For $s<s_c$ the saddle-point density is bounded while at $s=s_c$ the lower edge $\lambda_-(s_c,w)=0$ reaches the origin and $\rho_{s,w}$ acquires an inverse square root divergence there.

From $\rho_{s,w}$ the corresponding values of  scaled log-determinant $H$ and trace $T$ are
\be
\begin{cases}
\displaystyle h(s,w)= \int\de\rho_{s,w}(\lambda)\,\ln\lambda=\varphi\left(s-s_c\right)-\ln(1+2w) \\
\displaystyle t(s,w)= \int\de\rho_{s,w}(\lambda)\,\lambda=\displaystyle\frac{c-2s}{1+2w}\ ,
\end{cases}\label{eq:htstar}
\ee
 with $\varphi(x)=-1+(1-2x) \ln{\left(1-2x\right)}+2x \ln{(-2x)}$ for $x\leq0$  (hence $h(s,w)$ is defined for $s\leq s_c$). 
Combining~\eqref{eq:htstar} and~\eqref{eq:int_diff} we obtain the cumulant GF $J(s,w)$ as in Result~\ref{thm:A}.

In principle, one may also compute the rate function $\Psi(h,t)$ in the same way. However, it is not possible to write down $s(h,t)$ and $w(h,t)$ (the inverse maps of~\eqref{eq:htstar}) in terms of elementary functions. For simplicity, however, we show how to carry out the explicit computation for trace and log-determinant separately and establish the large $n$ decay as in Remark~\ref{rmk:4}. Setting $s=0$ we find $t(0,w)=c/(1+2w)$ from~\eqref{eq:htstar}, and we immediately get the rate function of the scaled traces from~\eqref{eq:int_diff} by integrating $w(t)$ (the inverse of $t(0,w)$)
\begin{align}
\Psi_T(t)=-\int_{t(0,0)}^t{w(t')\de t'}=\int_{c}^t{\frac{1}{2}\left(1-\frac{c}{t'}\right)\mathrm{d}t'}
=\frac{t-c}{2}+\frac{c}{2}\ln{\left(\frac{c}{t}\right)}.
\end{align}
This proves~\eqref{eq:traceratefunct}. Similarly, for the log-determinant $H$ we have
\be
\Psi_H(h)=-\int_{h(0,0)}^{h}s(h')\de h',\label{eq:rateH}
\ee
valid for $h>-1$ (see discussion in the next Section), where $s(h)$ is the inverse of $h(s,0)$.

\section{Further results and discussion}\label{sec:finite}

The treatment in the previous section does not cover the following two issues:
\begin{itemize}
\item The case $c=1$ (square data matrices), for which the leading term of the variance of $H=\ln\det\Ss$ (computed from the approach described above) is not defined (see Eq.~\eqref{varTH}). What is the origin of this hitch?
\item The origin of the condition $h>-1$ for the validity of the rate function $\Psi_H(h)$ in \eqref{eq:rateH} seems mysterious. What is the mechanism governing the statistics of ``anomalously small" log-determinants, then?
\end{itemize}
We discuss these two issues in detail here.

\subsection{The case $c=1$ (square data matrices)}

As already disclosed, if $\Ss$ is a Wishart matrix with $c=1$ ($m=n$) the limiting variance of $H=\ln\det\Ss$ is not described by our large deviations result (see Eq.~\eqref{varTH}). The origin of this hitch is as follows. Recall that the cumulant GF $J_H(s)$ is defined for $s\leq s_c=(c-1)/2$. 
Hence, for $c=1$ (i.e. $m=n$) $J_H(s)$ is non-analytic in $s=s_c=0$ and the cumulants cannot be obtained by differentiation. 

A way to circumvent this problem is to \emph{first} compute $\V(H)$ for \emph{finite} $n$, and then evaluate
its large $n$ asymptotics. The joint Laplace transform $\widehat{\mathcal{P}}_n(s,w)$ of $H$ and $T$ can be indeed evaluated exactly also at finite $n,m$, using the Laguerre-Selberg integral~\cite{Selberg42,forrester,Guionnet}: 
\be
\frac{1}{n!}\int_0^{+\infty}\hspace{-3mm}\cdots\int_0^{+\infty}\prod_{i< j}|x_i-x_j|^{2p}\prod_{i=1}^n x_i^{q-1} \e^{-x_i}\de x_i=\prod_{j=0}^{n-1}\frac{\Gamma(q+jp)\Gamma\left((j+1)p\right)}{\Gamma(p)},\qquad p,q>0\ .\label{eq:SelbergGuionnet}
\ee
Using this identity, one may evaluate the Laplace transform $\widehat{\mathcal{P}}_n(s,w)$ as
\be
\widehat{\mathcal{P}}_n(s,w)=\left(\frac{(\beta n/2)^{s }}{\left(2w+1\right)^{\frac{1}{2}+(\alpha-s)+\frac{1}{2n}}}\right)^{\beta n^2}\,\,\prod_{j=0}^{n-1}{\frac{\Gamma\left(\frac{\beta}{2} \left(j+2n(\alpha-s)\right)+1\right)}{\Gamma\left(\frac{\beta}{2}\left(j+2n\alpha\right)+1\right)}}\ , \label{laplexact}
\ee
with $\alpha$ as in \eqref{eq:potential_param}. 
We have verified that our large $n$ formulae reproduce with good accuracy the finite $n,m$ result even for moderate values of $n$.

From \eqref{laplexact} it is possible to extract the large deviation functions for the scaled trace $T$ (this corresponds to $s=0$). On the other hand, the asymptotic in the variable $s$ is not trivial.

\begin{figure}[t]
\centering
\includegraphics[width=.7\textwidth]{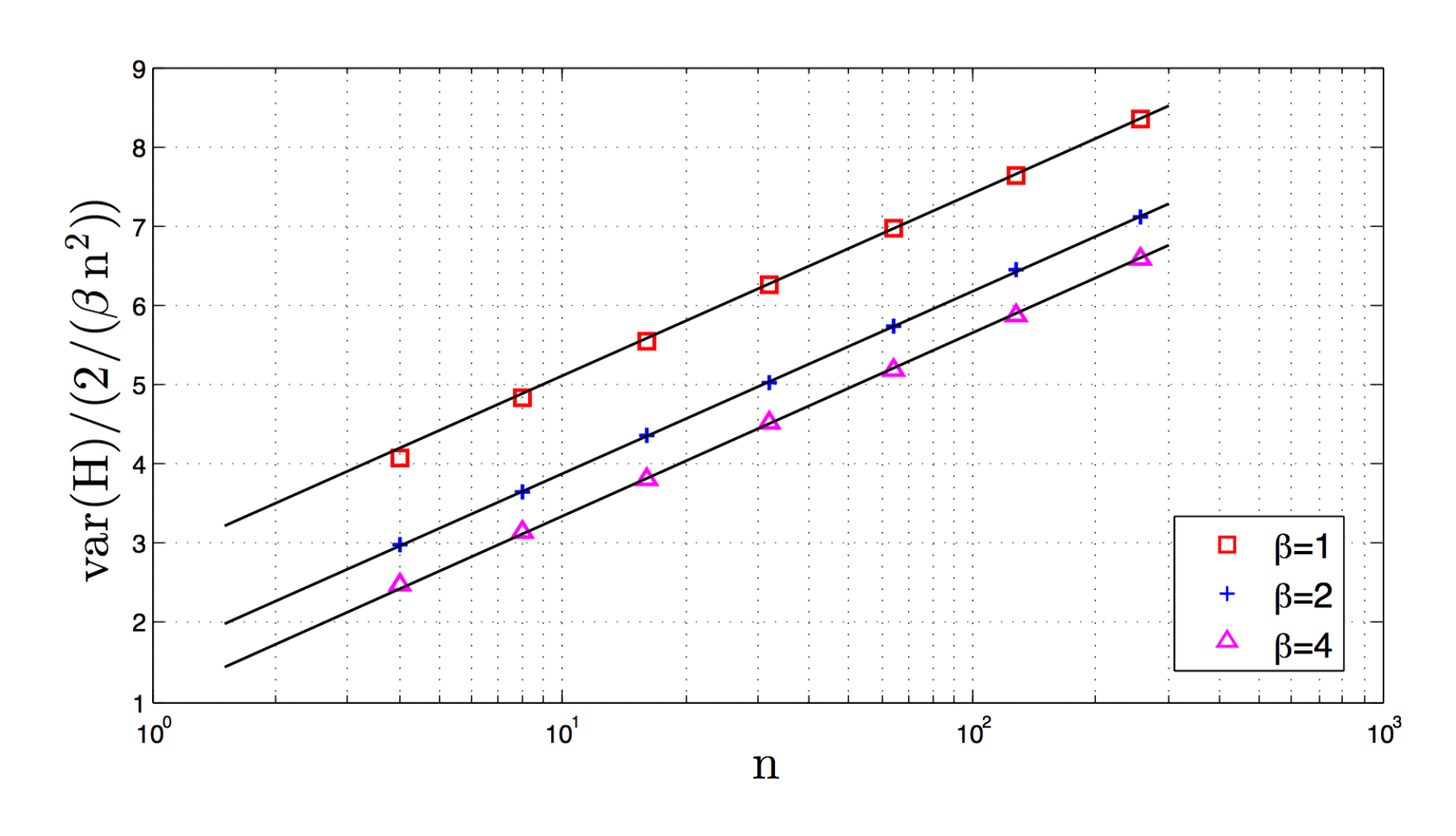}
\caption{Rescaled variance of the log-determinant $H=n^{-1}\sum_i\ln(\lambda_i)$ for $c=1$. 
Each point is produced sampling $N=10^6$ Wishart matrices of size $n$ for $\beta=1,2$ and $4$. The error for each point is of order $\Ord(10^{-2})$, not visible in the picture. The solid lines are the exact result~\eqref{eq:new_var_H}.}
\label{fig:log_var}
\end{figure}

However we can use this exact result to deduce $\V(H)$ for symmetric data matrices ($c=1$), the case that was not covered by our Result~\ref{thm:A}. Setting $c=1$ and $w=0$ in~\eqref{laplexact}, we denote by $\widehat{\mathcal{P}}_{n}(s)\equiv\widehat{\mathcal{P}}_n(s,0)=\langle \e^{-\beta n^2 s H}\rangle$ the Laplace transform of $H$ at finite $n=m$.
We can compute the derivatives of $\widehat{\mathcal{P}}_n(s)$ as 
\begin{align}
\widehat{P}_n'(s)&= \beta n^2\ln\left(\frac{\beta n}{2}\right)\ \widehat{P}_n(s)-\beta n \widehat{P}_n(s) \sum_{j=0}^{n-1}\psi_0\left(\frac{\beta}{2}(j+1-2 n s)\right)\ ,\\
\nonumber\widehat{P}_n''(s)&=\beta n^2\ln\left(\frac{\beta n}{2}\right) \widehat{P}_n'(s)-
\beta n \widehat{P}_n'(s)\sum_{j=0}^{n-1}\psi_0\left(\frac{\beta}{2}(j+1-2 n s)\right)\\
&+(\beta n)^2 \widehat{P}_n(s)\sum_{j=0}^{n-1}\psi_1\left(\frac{\beta}{2}(j+1-2 n s)\right)\ ,\label{eq:P2ndder}
\end{align}
where $\psi_m(z)=\partial_ z^{m+1}\ln\Gamma(z)$ is the $m$-Polygamma function~\cite{Abramowitz70}.  In principle one can compute higher derivatives recursively, and evaluate the asymptotic values of the cumulants of $H$. For instance, 
average and variance of $H$ are related to the derivatives $\widehat{P}_n'(s)$ and $\widehat{P}_n''(s)$ at $s=0$. Using the normalization $\widehat{P}_n(0)=1$ we get
\begin{align}
\langle H\rangle&=-\frac{1}{\beta n^2}\widehat{P}_n'(0)=-\ln\left(\frac{\beta n}{2}\right)+\frac{1}{n}\sum_{j=0}^{n-1}\psi_0\left(\frac{\beta}{2}(j+1)\right)\ .\label{eq:1}
\end{align}
Using the Euler-Maclaurin summation formula~\cite{Abramowitz70}
$\sum_{k=0}^{N}F(a+hk)=\frac{1}{h}\int_a^{b}\de t\,F(t)+\frac{1}{2}\left[F(b)+F(a)\right]+\dots$ (with $b=a+hN$),
and the classical asymptotic 
$\ln\Gamma(az+b)\sim\ln(\sqrt{2\pi})-az+\left(az+b-\frac{1}{2}\right)\ln(az)$,  valid for $z\to\infty$, with $|\arg z|<\pi$ and $a>0$,
we obtain for large $n$ the limit value  $\langle H\rangle\to-1$, according to~\eqref{averageTH} for $c=1$. A similar analysis of the Laguerre-Selberg integral (for $w=0$) was performed in~\cite{Cicuta}, but it was restricted to the computation of $\avg{H}$ at leading order in $n$. Here we tackle the problem of the variance of $H$ for square data matrices. From~\eqref{eq:P2ndder} we obtain
\begin{align}
\mathrm{var}(H)=\langle H^2\rangle-\langle H\rangle^2=\frac{1}{(\beta n^2)^2}\left\{\widehat{P}_n''(0)-\widehat{P}_n'(0)^2\right\}=\frac{1}{n^2}\sum_{j=0}^{n-1}\psi_1\left(\frac{\beta}{2}(j+1)\right)\ .\label{eq:2}
\end{align}
After a somewhat lengthy calculation, we managed to extract the large $n$ asymptotics 
\be
\frac{\V(H)}{\omega_{\beta}(n,2)}= \ln n+\ln(\beta/2)-\psi_0(\beta/2)+ 1+K_{\beta}+o(1)\ , \label{eq:new_var_H}
\ee 
with a constant $K_\beta$ given by 
\be
K_\beta=(2/\beta^2)\int_0^{\infty}t\,\e^{-2t(1-1/\beta)}(\beta \e^{2t/\beta}-2\e^t+2-\beta)(1-\e^{-t})^{-1}(\e^{2t/\beta}-1)^{-2}\de t\ . 
\ee  
Some special values are $K_1=\pi^2/8-\ln 2, K_2=0, K_4=1-\pi^2/8$. This result has been verified numerically, see Fig.~\ref{fig:log_var}. Note the logarithmic growth of~\eqref{eq:new_var_H} with $n$, in contrast to the $\Ord(1)$ limiting behavior of $\V(H)/\omega_{\beta}(n,2)$ for $c>1$. 
Such a logarithmic divergent variance is customary for \emph{discontinuous} spectral linear statistics in RMT, the paradigmatic example being the number variance~\cite{Dyson62,Majumdar11,Majumdar13,Marino16}. Notice that  the function $\ln \lambda$ is indeed discontinuous at $\lambda=0$. However, as long as $s<s_c$  the support of the equilibrium measure $\rho_{s,w}(\lambda)$ does not contain the origin and this singularity is ineffective; only for $s=s_c$ we have $\lambda_-(s_c,w)=0$, and at that point the singularity of $\ln\lambda$ starts being felt. The central limit theorem with logarithmically divergent variance for $H$ has been proved in~\cite{Rempala05,Nguyen14} for $\beta=1$. The subleading corrections to $\V(H)$ in~\eqref{eq:new_var_H} are instead a new result.

\subsection{The statistics of atypically small log-determinants}\label{small}

We have claimed earlier that the rate function $\Psi_H(h)$ of $H$ can be computed as the inverse Legendre-Fenchel transform of $J_H(s)$ only for $h>-1$. Why is this the case? As a matter of fact, the G\"artner-Ellis theorem has two hypotheses: first, the cumulant GF is required to be differentiable in the interior of its domain; second, the derivatives of  the cumulant GF should diverge on the boundaries of the domain (a condition known as \emph{steepness}~\cite{Ellis84,Touchette09}). In our case, $J_H(s)$ is differentiable for all $s<s_c$ but the left derivative attains a \emph{finite} value at the boundary point $s_c$: $\partial_sJ_H(s)\to-1$ as $s\to s_c^-$. Hence, only a \emph{local} version of the G\"artner-Ellis theorem holds and  $\PP_H(h)\approx \exp(-\beta n^2 \Psi_H(h))$ with $\Psi_H(h)$ being the branch of the Legendre-Fenchel transform of $J_H(s)$ for $h>-1$. 

The 2D Coulomb gas analogy provides a rather intuitive physical picture of this obstruction. We have seen that the saddle-point density $\rho_{s,w}$ is bounded as long as $s<s_c$ (see~\eqref{rhostar}-\eqref{eq:support}). When $s=s_c$, the lower edge of the saddle-point density reaches the origin $\lambda_-(s_c,w)=0$ and $\rho_{s_c,w}(\lambda)\sim\lambda^{-1/2}$ acquires an inverse square-root singularity there. For $s>s_c$, the logarithmic part of the effective potential $V(\lambda)+U_{s,w}(\lambda)$ becomes \emph{attractive}, giving rise to an electrostatic instability of the gas. As already discussed, $\rho_{s,w}(\lambda)$ is the typical distribution of the eigenvalues of $\Ss$ yielding a prescribed value of $H=\int\de\rho_{s,w}(\lambda)\,\ln\lambda$. Setting $w=\alpha=0$ to simplify the discussion, we see that $\int\de\rho_{s,0}(\lambda)\,\ln\lambda>-1$ as long as $s< s_c$ and the critical value $H_{\text{cr}}=-1$ corresponds to the critical density $\rho_{s_c,0}(\lambda)=\frac{1}{2\pi}\sqrt{(4-\lambda)/\lambda}\,{\bm 1}_{\lambda\in(0,4)}$ obeying $\int\de\rho_{s_c,0}(\lambda)\,\ln\lambda=H_{\text{cr}}$. A solution to the problem of smaller log-determinant $H<H_{\text{cr}}$ would be achieved if the typical distribution of the eigenvalues corresponding to this anomalously small $H$ were known. 

What is then the behavior of the Coulomb gas constrained to have $H<H_{\text{cr}}$? As suggested in~\cite{Cunden_unp}, a failure of the steepness condition may be the hallmark of split-off phenomena of random variables. Guided by numerics and intuition, since the function $\ln\lambda$ is divergent for $\lambda\downarrow0$, we expect that atypically small values of $H=n^{-1}\sum_{i=1}^n\ln\lambda_i<H_{\text{cr}}$ are driven by the statistical behavior of the smallest eigenvalue $\lambda_{\min}$.  For $H>H_{\text{cr}}$ the Coulomb gas particles behave `cooperatively' to accommodate atypical values of $H$ (each of the random variables $\ln\lambda_i$'s contributes to realize $H$). On the contrary, large fluctuations of $H<H_{\text{cr}}$ are typically realized by fluctuations of $\lambda_{\min}$ to the left (the random variable $\ln\lambda_{\min}$ contributes macroscopically to $H$). 
This line of reasoning would imply a change of scaling (speed) in the large $n$ behavior of the probability density of $H$.
The idea is to split the contribution of the Coulomb gas to $H$ in two parts:
\be
H[\rho]=\frac{1}{n}\ln\lambda_{\mathrm{min}}+H[\tilde{\rho}]\ ,\quad\text{where}\quad \tilde{\rho}(\lambda)=\frac{1}{n}\sum_{i:\lambda_i\neq\lambda_{\mathrm{min}}}\delta(\lambda-\lambda_i)\ .
\ee
The probability density of $H$ can be written as
\be
\PP_{H}(h)=\int_{0}^{+\infty}\de x\mathcal{P}_H\left(h\,|\,\lambda_{\mathrm{min}}=x\right)\PP_{\lambda_{\mathrm{min}}}(x)\ .\label{eq:logdet_cond}
\ee
At this point we need to understand the distribution of the smallest eigenvalue $\PP_{\lambda_{\mathrm{min}}}(x)$ and the conditional probability $\mathcal{P}_H\left(h\,|\,\lambda_{\mathrm{min}}=x\right)$.
It is easy to show that the probability density function of the smallest eigenvalue behaves for large $n$ as
\be
\PP_{\lambda_{\mathrm{min}}}(x)\approx \mathrm{e}^{-\beta n^2 x/2}\ ,
\ee
corresponding to a typical value $\avg{\lambda_{\mathrm{min}}}=2/(\beta n^2)$ and   $\V(\lambda_{\mathrm{min}})=4/(\beta^2 n^4)$ at leading order in $n$.
Typical fluctuations of order $\Ord(n^{-2})$ to the right of $\avg{\lambda_{\mathrm{min}}}$ are irrelevant for the statistical behavior of $H$. On the contrary the typical fluctuations to the left ($\lambda_{\mathrm{min}}<\avg{\lambda_{\mathrm{min}}}$) play a significant role due to the divergent character of $\ln\lambda_{\mathrm{min}}$ for $\lambda_{\mathrm{min}}\downarrow0$. Roughly speaking, the \emph{typical} fluctuations of order $\Ord(n^{-2})$ of the smallest eigenvalues do not change the limiting macroscopic density of the eigenvalues ($\tilde{\rho}(\lambda)\simeq \rho_{s_c,0}(\lambda)$, irrespective of the value of $H\leq H_{\mathrm{crit}}$), but nevertheless have dramatic consequences on the statistics of $H$. 

Similar evaporation phenomena for both correlated and i.i.d. variables have been recently detected in a variety of contexts (see e.g. \cite{DePasquale10,Texier13,Filiasi14,Evans14,Vivo07,Burda07,Facchi13}). The new interesting twist here is that the split-off is realized by the \emph{smallest} (and not the usual \emph{largest}) of the random variables involved.
Here, using $H\left[\tilde{\rho}\right]\simeq H\left[{\rho_{s_c,0}}\right]=-1$, for $0\leq x\leq \avg{\lambda_{\mathrm{min}}}$ we have:
\begin{align}
\mathcal{P}_H\left(h\,|\,\lambda_{\mathrm{min}}=x\right)
&=\mathcal{P}_H\left(h=\frac{1}{n}\ln\lambda_{\mathrm{min}}+H\left[\tilde{\rho}\right]\,\bigg|\,\lambda_{\mathrm{min}}=x\right)
&\rightarrow
\delta\left(h-\left(\frac{1}{n}\ln\lambda_{\mathrm{min}}-1\right)\right)\ .
\end{align}
Using the above result, from \eqref{eq:logdet_cond}, we easily get
\be
\PP_{H}(h)\approx \mathrm{e}^{-n\tilde{\psi}_H(h)},\quad\text{with}\quad\tilde{\psi}_H(h)=-1-h\ , \label{eq:rateHsmall}
\ee
for $h<-1$ (note the speed $n$ in contrast to the speed $n^2$ in the ``democratic" Coulomb gas setting). 

A further argument in support of this change of speed can be obtained for $\beta=2$ and $\alpha=0$ using a finite-$n$ approach based on 
the Laguerre-Selberg integral~\eqref{eq:SelbergGuionnet}. 
It is convenient to work directly at the level of the determinant of $\mathcal{S}$. 
Let $\mathcal{P}_{\hat{G}}(\hat g)$ be the probability density of the determinant $\hat G=\det(\mathcal{S})=\prod_{i=1}^n\lambda_i$  (without the power $1/n$). 
Its Mellin transform is given by

\be
\hat{M}(\hat s)=\int_0^\infty \de\hat g \mathcal{P}_{\hat G}(\hat g)\hat g^{\hat{s}-1}=
 \left(\frac{1}{n}\right)^{n (\hat s-1)} \frac{G(n+\hat s)\Gamma(\hat s)}{G(n+1)G(1+\hat s)}\ ,
\label{mellin1}
\ee
where $G(x)$ is the Barnes G-function. Using the asymptotics~\cite{Abramowitz70}
\be
\frac{G(n+\hat s)}{G(n+1)}\sim \exp\left[n (1-\hat s) -n\ln n +\hat s\ n \ln n +\ln n [1/2-\hat s+\hat s^2/2]  -\ln(2\pi) (1-\hat s)/2 +o(1/n)\right],
\ee
valid  for $n\to\infty$, we obtain (with logarithmic accuracy)
\be
\hat{M}(\hat s)\approx \e^{-n(\hat{s}-1)}\ .\label{mellinexp}
\ee
This Mellin transform can be written also in terms of the probability density  $\mathcal{P}_H(h)$ of $H=n^{-1}\ln\hat{G}$. Assuming  $\mathcal{P}_H(h)\approx \e^{-n \tilde{\psi}_H(h)}$ for $h<-1$ we have
\be
\hat{M}(\hat s)=\int_{-\infty}^\infty \de h \PP_H(h)e^{n(\hat{s}-1)h}\approx \int_{-\infty}^{-1} \de h\ \e^{-n\left[\tilde{\psi}_H(h)-(\hat{s}-1)h\right]}\approx \exp\left[-n\min_h \left(\tilde{\psi}_H(h)-(\hat{s}-1)h\right) \right]\ .\label{mellinexp2}
\ee
Here the integral has been truncated at $h=-1$ since $\PP_H(h)$ decays faster (exponentially with speed $n^2$) for $h>-1$, and  Laplace's approximation has been used in the last step. Matching \eqref{mellinexp} with \eqref{mellinexp2}, we eventually obtain $\tilde{\psi}_H(h)$ as in~\eqref{eq:rateHsmall}.

\section{Conclusions}
\label{sec:conclusions}
In summary, we have considered the joint statistics (including large deviation tails) of generalized and total variance of a large $n\times m$ Gaussian dataset. These observables are just the scaled log-determinant $H$ and the trace $T$ of the corresponding $n\times n$ covariance matrix. We have employed a powerful combination of two techniques: the Coulomb gas analogy of statistical physics, which allowed us to represent the eigenvalues of the covariance matrix as an interacting gas of charged particles, whose excess free energy is the the cumulant generating function for our observables in the limit $n,m\to \infty$ with $c=m/n$ fixed, and a finite $n,m$ approach based on the Laguerre-Selberg integral. Combining these two approaches, we complemented the Coulomb gas method with two interesting cases that fell out of its domain: i) the case $c=1$ (square datasets), for which the excess free energy is non-analytic in zero. This has the consequence that the variance of H grows logarithmically with $n$, with a subleading constant term that we could precisely characterize, and ii) atypically small log-determinants, for which the corresponding rate function in Laplace space is non-steep. This implies an abrupt change of speed in the corresponding large deviation principle, which can be ascribed to the \emph{split-off} of the smallest eigenvalue from the unperturbed Mar\v{c}enko-Pastur distribution. This picture is supported by numerical simulations and a saddle-point argument based on a finite $n,m$ formula (see Subsection \ref{small}). 

It would be interesting to investigate whether our results could be extended to non-Gaussian and possibly correlated data matrices.  Our derivation strongly relied on the data being normally distributed and a different approach seems to be needed for more general covariance matrices.

\section*{Acknowledgments}
FDC acknowledges support by EPSRC Grant number EP/L010305/1 and  partial support of Gruppo Nazionale di Fisica Matematica GNFM-INdAM.
PV acknowledges the stimulating research environment provided by the EPSRC Centre for Doctoral
Training in Cross-Disciplinary Approaches to Non-Equilibrium Systems (CANES,
EP/L015854/1). FDC acknowledges hospitality by the LPTMS (Univ. Paris Sud) where this work was initiated. The results presented in this paper are part of the first author's Ph.D. thesis at the University of Bari, Italy.  The authors thank P. Facchi, A. Lerario and M. Marsili for helpful advices and very stimulating discussions. FDC thanks S. Di Martino for technical support.

\end{document}